\documentclass[aps,prd,superscriptaddress,amsfonts,amssymb,amsmath,eqsecnum,nofootinbib,twocolumn,floatfix]{revtex4}
 \usepackage{color,graphicx}
 \usepackage[utf8]{inputenc}
 \usepackage[english]{babel}
 \definecolor{darkblue}{rgb}{0,0,0.7}
\definecolor{darkred}{rgb}{0.7,0,0}
\definecolor{darkgreen}{rgb}{0,0.4,0}
\usepackage[unicode, colorlinks, citecolor=darkblue, linkcolor=darkred, urlcolor=blue]{hyperref}

\allowdisplaybreaks
\begin{document}

\author{Sergey P. Vyatchanin}
\affiliation{Faculty of Physics, M.V. Lomonosov Moscow State University, Leninskie Gory, Moscow 119991,  Russia}
\author{Andrey B. Matsko}
\affiliation{ OEwaves Inc., 465 North Halstead Street, Suite 140, Pasadena, CA 91107}

\date{\today}
	
\title{On sensitivity limitations of a dichromatic optical detection\\ of a classical mechanical force}

\begin{abstract}
We apply the strategy of the back action evading measurement of a quadrature component of mechanical motion of a test mass to detection of a classical force acting on the mass  \cite{80a1BrThVo} and study both classical and quantum limitations of the technique. We are considering a resonant displacement transducer interrogated with a dichromatic optical pump as a model system in this study. The transducer is represented by a Fabry Perot cavity with a totally reflecting movable end mirror the resonant force of interest acts upon. The cavity is pumped with two coherent optical carriers equally detuned from one of the cavity resonances. We show that the quantum back action cannot be completely excluded from the measurement result due to the dynamic instability of the opto-mechanical system that either limits the allowable power of the optical pump or calls for introducing an asymmetry to the pump configuration destroying the quantum nondemolition nature of the measurement.
\end{abstract}

\maketitle

\section{Introduction}

Opto-mechanical systems \cite{Marquardt09,AspelmeyerRMP2014} are utilized in high precision measurements including gravitational wave detection \cite{aLIGO2013,aLIGO2014,aLIGO2015}, torque sensing \cite{WuPRX2014}, magnetometery \cite{ForstnerPRL2012}, and intracavity photon number as well as optical quadrature component detection \cite{94a1BrGoKh,03a1BrGoKhMaThVy}. The maximum sensitivity of the measurements is fundamentally limited because of the quantum noise as well as quantum uncertainty associated with the optical as well as mechanical degrees of freedom. Namely, uncertainty of the initial coordinate and momentum of the mechanical degree of freedom limits the accuracy of the measurement of the mechanical coordinate. The quantum noise of the optical probe adds to the measurement error directly via transfer to the classical meter noise as well as via quantum back action disturbing the mechanical system \cite{BrKh92,02a1KiLeMaThVy, 03a1BrGoKhMaThVy}. The impact of the quantum noise can be reduced by
selecting an optimal measurement strategy. For instance, it is possible to exclude the effect of the quantum uncertainty of initial conditions on the measurement sensitivity in the case of classical force detection \cite{BrKh92,03a1BrGoKhMaThVy}. It is also possible to avoid quantum back action in the variational measurements \cite{02a1KiLeMaThVy}. As such, the number of known back action avoiding measurement techniques is limited and it is fundamentally important not only to find strategies resulting in the increase of the measurement sensitivity beyond the boundaries introduced by the quantum effects but also to understand fundamental restrictions of the existing strategies. In this paper we adopt the technique of back action evading measurement of a quadrature component of a mechanical oscillator originally proposed by Vladimir Braginsky and colleagues \cite{80a1BrThVo} to optical detection of a resonant force acting on a mechanical oscillator and find limitations of the method.

The sensitivity of the continuous measurement of a mechanical coordinate is restricted by so called standard quantum limit (SQL) appearing due to the quantum back action  \cite{Braginsky68,BrKh92}. Increase of the power of the optical pump allows improving the sensitivity associated with the optical shot noise because of increase of the signal-to-noise ratio of the measurements. The quantum back action results from the disturbance of the coordinate due to fluctuations of the light pressure coming from the fluctuations of photon number in the optical cavity. Because of this process the sensitivity of the coordinate measurement improves up to a point with the optical power increase and than drops. There exists an optimal power value that results in the maximum measurement sensitivity in this case. The SQL was studied in various configurations ranging from macroscopic kilometer-sized gravitational wave detectors \cite{02a1KiLeMaThVy} to microcavities \cite{Kippenberg08,DobrindtPRL2010}.

Detection of a classical force acting on a suspended test mass is an example of a coordinate measurement and, hence, there exist an SQL of the force detection. In this case, though, one measures not the absolute coordinate of  the mass, but its modification due to the force action. If the envelope of the force of interest is known, the impact of the initial uncertainty of the coordinate and momentum of the probe mass on the measurement sensitivity can be suppressed \cite{BrKh92,03a1BrGoKhMaThVy}. The SQL of the force measurements can be avoided with several approaches including
stroboscopic position measurement \cite{78a1eBrKhVo}, mechanical quadrature measurement \cite{78ThornePRL}, variational measurement \cite{93a1VyMaJETP,95a1VyZuPLA, 02a1KiLeMaThVy}, opto-mechanical velocity measurement \cite{90BrKhPLA,00a1BrGoKhThPRD,16a1PRAVyMa}, and measurements performed in opto-mechanical systems with ponderomotive rigidity \cite{99a1BrKhPLA, 01a1KhPLA}.
\begin{figure}[b]
\includegraphics[width=0.47\textwidth]{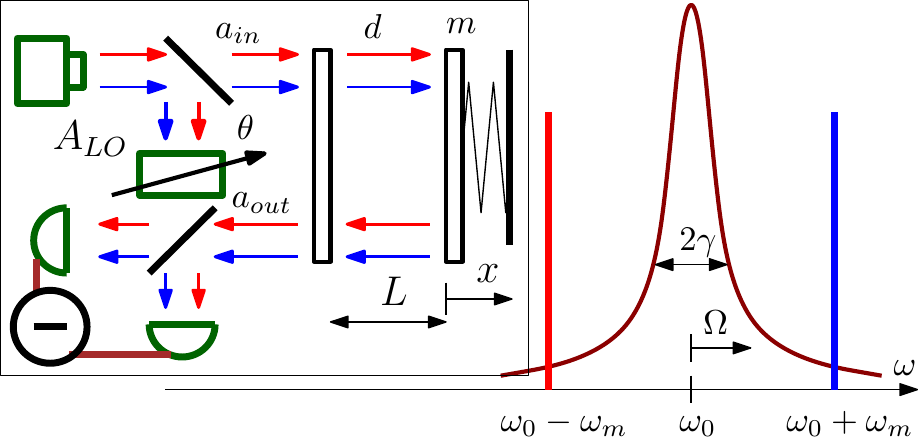}
\caption{The Fabry-Perot (FP) optical cavity with suspended back mirror is pumped by light consisting of two frequency harmonics. The carrier frequencies of the light are tuned symmetrically with respect to the eigen frequency $\omega_0$ of the optical mode of interest: the red detuned carrier with   frequency $\omega_-=\omega_0-\omega_m$ and blue detuned carrier (frequency $\omega_+= \omega_0+\omega_m $). Output wave is detected by balanced homodyne detector with local oscillator taken from pump delayed by angle $\theta$.}\label{Dpump2}
\end{figure}

In this paper we introduce a measurement technique suitable for the detection of a resonant force. We show theoretically a possibility of surpassing the SQL in a classical force measurement involving a Fabry-Perot (FP) cavity with the end mirror represented by a suspended test mass of mechanical oscillator. The force of interest acting on the test mass is nearly resonant with the mechanical oscillator. The optical cavity is pumped with light consisting of two frequency harmonics {\em equally} detuned from the resonance frequency of a cavity mode to blue and red,  as illustrated in Fig.~\ref{Dpump2}.

It was shown by Braginsky, Vorontsov, and Thorne that a modulated RF pump can be used to perform a QND measurement of a quadrature component of a mechanical system that can be utilized for a classical force detection \cite{80a1BrThVo}. The idea was expanded to the optical domain and possibility of a QND measurement of a quadrature component of a mechanical degree of freedom in an opto-mechanical system was confirmed \cite{Clerk08} and further applied to conditional squeezing \cite{Clerk2015, Clerk2016}.

We here extend the idea and show that the QND is applicable for the classical force detection. Similarly to the previous works we have found that the quantum back action can be completely avoided when i) the power values of the harmonics are identical and ii) one observes the optical harmonics in the vicinity of the resonance of the optical cavity mode. The red detuned pump introduces positive damping and can be used for cooling the mechanical oscillator, while the blue-detuned pump results in the gain and may lead to instability of the system and opto-mechanical oscillations \cite{Marquardt09,AspelmeyerRMP2014}. The damping compensates the gain, so the system remains stable up to a point. It was proposed to use a monochromatic optical local oscillator to measure the mechanical quadrature amplitude \cite{Clerk08}. We propose to use a dichromatic local oscillator in the measurement. This is more natural as we can utilize the phase shifted pump light to perform such a measurement. This technique can be
called as a synodyne (in sense \cite{Buchmann2016}) QND measurement.

We study the limitation of the QND technique and find that the stability range of the opto-mechanical system is compromised due to its salient nonlinear properties. The nonlinearity comes not because of an avoidable nonlinearity of the mechanical system itself, but from the opto-mechanical interaction \cite{96a1VyMaZu}. We found that it limits the dynamic stability of both the QND measurement of quadrature discussed in \cite{Clerk08} and the sensitivity of force measurement, proposed in this paper. The system starts oscillating at some power threshold. Up to our knowledge, this type of an opto-mechanical instability was not discussed previously.

\section{Linear Analysis of the Force Detector}

Let us consider an interferometric displacement transducer consisting of a Fabry-Perot (FP) cavity of length $L$ and an ideal movable end mirror suspended on a single dimension mechanical oscillator (as shown in Fig.~\ref{Dpump2}). A cavity mode characterized with eigen frequency $\omega_0$ and relaxation rate $\gamma$ (half width at the half maximum) resulting from the partially transparent front mirror is interrogated with a bichromatic coherent optical pump. The expectation value of the amplitude of the optical pump is presented in form
\begin{align}
\label{mathA}
\mathcal A(t) =  A_+e^{-i(\omega_0 +\omega_m) t} + A_-e^{-i(\omega_0-\omega_m) t},
\end{align}
where $A_+$ ($A_-$) is complex amplitude of pump detuned to  blue (red) side from resonance frequency of cavity, see also \eqref{alphaIn} for introduction of the amplitude in the quantum case.

The mirror is characterized with coordinate
\begin{align}
  \label{x(t)}
 x(t)=x_z\left(\hat b^\dag  + \hat b \right),\quad x_z\equiv \sqrt\frac{\hslash}{2m\omega_m}
\end{align}
where $\hat b^\dag,\, \hat b$ are the creation and annihilation operators, $x_z$ is the amplitude of zero-point mechanical fluctuations, $m$ is the mass, $\omega_m$ is the mechanical eigenfrequency, $\gamma_m$ is the mechanical relaxation rate. The resonant classical force of interest, $F(t) = F_s(t) \cos \omega_m t$, acts upon the mirror, where $F_s$ is a slow amplitude of the force.

\subsection{Major equations}

We assume that resolved side band conditions
\begin{align}
  \label{rsb}
 \omega_m \gg \gamma\gg \gamma_m
\end{align}
are valid and utilize Hamiltonian
\begin{subequations}
\label{Hamiltonian}
 \begin{align}
   H& =\hbar \omega_0 \left ( 1-\frac{x}{L} \right )\hat d_i^\dag \hat d_i +H_\gamma
   + \hslash\omega_m \hat b^\dag \hat b + \\
   \label{H1}
    & +H_{\gamma_m}  - F(t)\, x_z \left(\hat b + \hat b^\dag\right), \\
H_\gamma &= i\hbar  \sqrt{2 \gamma}\,
    \left [\hat d_i^\dag \hat \alpha_\text{in} -\hat \alpha_\text{in}^\dag \hat d_i    \right ]
    +H_{\gamma\, bath}, \\
H_{\gamma_m}&= i\hbar  \sqrt{2 \gamma_m}\, \left [ \hat b^\dag \hat \beta _\text{th} - \hat b \hat \beta^\dag_\text{th}   \right ] + H_{\gamma_m\, bath},
 \end{align}
\end{subequations}
where $\hat d_i$ and $\hat d^\dag_i$ are annihilation and creation operators describing the intracavity optical field, $\hat \alpha_\text{in}$ is an operator of the external field,
last term in \eqref{H1} describes action of the signal force,
$H_\gamma$ describes attenuation of the pump photons and associated quantum noise,  $H_{\gamma_m}$ stands for the dissipation in the mechanical oscillator due to interaction with the thermal bath, $H_{\gamma\, bath}, H_{\gamma_m\, bath}$ represent Hamiltonians of thermal bathes (optical and mechanical, correspondingly) \cite{Marquardt09, AspelmeyerRMP2014}, see also details of the Hamiltonian approach in \cite{Law1995, Dalvit2000}.

From \eqref{Hamiltonian} we obtain equations of motion
\begin{subequations}
 \label{initEq}
 \begin{align}
 \label{initEq1}
  \partial_t \hat d_i +\left[\gamma+i\omega_0\left(1-\frac{x_z\left[ \hat b +\hat b^\dag\right]}{L}\right)\right] &\hat d_i
  =\\
  &= \sqrt{2\gamma} \hat \alpha_\text{in},\nonumber\\
  \label{initEq2}
  \partial_t \hat b +(\gamma_m+i\omega_m)\hat b -\frac{i\omega_0 x_z}{L}\, \hat d_i^\dag \hat d_i&= \\
    =\sqrt{2\gamma_m}\hat \beta_\text{th} +& \frac{iF_s e^{-i\omega_m t}}{2\sqrt{2\hslash m\omega_m}},\nonumber
 \end{align}
 which should be supplied by equation for output amplitude $\hat  \alpha_\text{out}$
 \begin{align}
  \hat  \alpha_\text{out} &= -  \hat \alpha_\text{in}+\sqrt{2 \gamma } \hat d_i\,.
 \end{align}
\end{subequations}

The field of opto-mechanics has established certain conventions over the last decade (see \cite{AspelmeyerRMP2014} for details). Our notations, primarily belonging to the analysis of optical force detectors \cite{02a1KiLeMaThVy}, have direct relations with those. For example, half width at the half maximum $\gamma$ for the optical cavity is usually denoted as $\kappa/2$ in opto-mechanics literature; the opto-mechanical coupling strength is usually expressed in frequency terms, i.e. $g = x_z \omega_0/L$.

The model \eqref{initEq} is similar but not equivalent to the adiabatic treatment of conservative cavity with slow mirror movement discussed in \cite{Law1995} in case of zero relaxation rate of cavity ($\gamma=0$). It is necessary to take into account not only the displacement of the mirror but also its velocity (not accounted in \eqref{initEq1}) for the precise consideration of adiabatic treatment in spirit of \cite{Law1995}.

The operators of the input, $\hat \alpha_\text{in}$, and the output, $\hat \alpha_\text{put}$, fields are presented as a sum of expectation, $\mathcal A$(t), and fluctuation, $a$, parts:
\begin{align}
\label{alphaIn}
 \hat \alpha_\text{in} &= \mathcal A(t)
	  +e^{-i\omega_0 t} \int_{-\omega_0}^\infty a_\text{in}(\Omega)\, e^{-i\Omega t}
	\frac{d\Omega}{2\pi}, \\
\hat \alpha_\text{out} &=\mathcal A_\text{out}(t) + \int_{-\omega_0}^\infty a_\text{out}(\Omega)\, e^{-i\Omega t}	\frac{d\Omega}{2\pi}\,,\\
 \hat \beta_\text{th} &= e^{-i\omega_m t}\int_{-\omega_m}^\infty b_\text{th}(\Omega)e^{-i\Omega t} \frac{d\Omega}{2\pi}\,
\end{align}
where $a_\text{in}(\Omega)$ describe vacuum fluctuations of light and obey to relationship for the commutators and correlators:
\begin{align}
  \label{comm1}
\big[a_\text{in}(\Omega),a_\text{in}^\dag(\Omega')\big] &= 2\pi \,\delta (\Omega-\Omega'), \\
\langle a_\text{in}(\Omega) a_\text{in}^\dag(\Omega') \rangle & =2\pi \,\delta(\Omega-\Omega').
\end{align}
The operator of the thermal bath $b_\text{th}$ obeys to the rules
\begin{align}
  \label{comm2}
 \big[b_\text{th}(\Omega),\, b^\dag_\text{th}(\Omega')\big] &= 2\pi \,\delta (\Omega-\Omega'),\\
 \langle b^\dag_\text{th}(\Omega) b_\text{th}(\Omega') \rangle &=2\pi \, n_T\,\delta(\Omega-\Omega'),
\end{align}
$n_T$ is  the number of the thermal quanta. The operators $\hat b$ as well as $\hat d_i$ are presented in a similar way
\begin{align}
  \label{bm}
   \hat b(t) & = e^{-i\omega_m t}\int_{-\omega_m}^\infty b (\Omega) \, e^{-i\Omega t} \,  \frac{d\Omega}{2\pi},\quad
       \\
    \hat d_i(t)& = \mathcal D(t) + e^{-i\omega_0 t}\int_{-\omega_0}^\infty d_i(\Omega) \, e^{-i\Omega t} \,\frac{d\Omega}{2\pi} \, ,\\
   \label{D(t)}
  \mathcal D(t) &= D_+e^{-i(\omega_0 +\omega_m) t} + D_-e^{-i(\omega_0-\omega_m) t},\\
  \label{Dpm}
  D_\pm &=\frac{\sqrt{2\gamma}}{\gamma \mp i\omega_m}\,A_\pm ,
\end{align}
where  $D_\pm$ are the mean amplitudes of the light circulating in the FP cavity at the pump frequencies.

It worth noting that we used the following rules to indicate the Fourier amplitudes through the manuscript. For all the operators in the paper the Fourier transform is denoted by dropping the hat and using the corresponding English alphabet letter. For instance,
\begin{align}
   \label{notations}
   b &= b(\Omega),\ b^\dag_-=b^\dag(-\Omega),
 \end{align}
The notations for the other variables are defined in a similar way.

The notations \eqref{notations} and commutation relations (\ref{comm1}, \ref{comm2}) are selected to be as in \cite{02a1KiLeMaThVy} which differ from ones used in \cite{AspelmeyerRMP2014}. As the result, in the opto-mechanic approach  \cite{AspelmeyerRMP2014} the Fourier-amplitudes $a(\Omega)= \int dt\, e^{i\Omega t} a(t)$ coincides with our expression \cite{02a1KiLeMaThVy}, whereas $a^\dag(\Omega)$ is defined as  $a^\dag(\Omega) = \int dt\, e^{i\Omega t} a^\dag (t) $, which is different from our definition $a^\dag(\Omega) = \int dt\,e^{-i\Omega t} a^\dag (t) $.

Our convention leads to correlators of the form $[a(\Omega),\,a^\dag(\Omega') = 2\pi \delta(\Omega - \Omega')$, while the "optomechanics convention" leads to $[a(\Omega),\,a^\dag(\Omega') = 2\pi \delta(\Omega + \Omega')$. The "optomechanics convention" requires to keep in mind that complex conjugation $a ^\dag (\Omega)$ of the modes equation comes with a sign change for the frequency: $a^\dag(\Omega) =(a(-\Omega))^\dag$, while our description calls for $a^\dag (\Omega) = (a (\Omega))^\dag$. It is easy to check that independently of the selected convention the result of the calculations does not change.

Linearising the set \eqref{initEq} in the vicinity of the steady state solution we rewrite it for Fourier transforms
\begin{subequations}
 \label{initEqFT}
 \begin{align}
 \label{initEqFT1}
   \left[\gamma -i\Omega\right] d_i   &= \sqrt{2\gamma} \hat a_\text{in}
     +  \frac{i\omega_0 x_z}{L}\left[ D_- b + D_+ b_-^\dag\right] ,\\
  \label{initEqFT2}
   \left(\gamma_m-i\Omega\right)b  &= \frac{i\omega_0 x_z}{L}\, \left[D_-^* d_i + D_+ d_{i-}^\dag\right] +\\
   &\qquad  +\sqrt{2\gamma_m} b_\text{th} + f_s, \nonumber\\
   \label{fs}
    & \quad f_s  \equiv \frac{iF_s(\Omega)}{2\sqrt{2\hslash m \omega_m}},\\
  \label{initEqFT3}
   a_\text{out} &= -  a_\text{in}+\sqrt{2 \gamma }  d_i\,.
 \end{align}
\end{subequations}
where $F_s(\Omega)$ is the  Fourier Transform of slow amplitude $F_s(t)$ of the resonance signal force.

\subsection{Solution}

To explain the proposed measurement technique we first consider an ideal case of the light harmonics equally detuned from the optical mode of interest with the detuning value equal to the mechanical frequency, see \eqref{D(t)}. Substituting \eqref{initEqFT1} into \eqref{initEqFT3} we get for the fluctuation amplitude of the output light  $a_\text{out}$
\begin{align}
\label{aout}
a_\text{out}&  =   \frac{\gamma+i\Omega }{\gamma-i\Omega }\, a_\text{in} %
	+ \frac{i \omega_0 x_z \sqrt{2\gamma}\left[ D_- b + D_+ b^\dag_-\right] }{L(\gamma - i \Omega) }\,.
\end{align}

The operators describing the mechanical oscillator can be presented in form
\begin{subequations}
\label{initbx}
\begin{align}
   \label{initbx1}
      b &= \frac{\sqrt{2\gamma_m}\,b_\text{th} + f_{s}}{\gamma_m +\Gamma -i\Omega}+\\
   \label{initbx2}
     &\quad + \frac{i \omega_0 x_z \sqrt{2\gamma}\left(D_-^*a_\text{in} +D_+ a^\dag_\text{in--}\right)}{L(\gamma -i\Omega)(\gamma_m +\Gamma -i\Omega)}\, ,\\
      \label{tildegamma}
     \Gamma &=
     \frac{(\omega_0 x_z)^2}{L^2(\gamma -i\Omega) }\left[|D_-|^2 -|D_+|^2\right]\,.
  \end{align}
  \end{subequations}
$\Gamma$ is the opto-mechanical damping parameter (dynamic back action). The term \eqref{initbx2} describes fluctuations due to the quantum back action.

Substituting \eqref{initbx} into \eqref{aout} we obtain
\begin{subequations}
\label{aoutfin}
\begin{align}
\label{aout1}
  a&_\text{out}  = e^{2i\eta}\cdot
      \frac{\gamma_m - \Gamma^* -i\Omega}{\gamma_m + \Gamma -i\Omega}\cdot a_\text{in} +\\
   +& 	 \frac{ i \sqrt{2G}\,e^{i\eta}}{( \gamma_m + \Gamma -i\Omega)}\left[\sqrt{2\gamma_m}\, d_\text{th}
	 + \frac{D_-f_s + D_+f_{s-}^*}{\sqrt{|D_+|^2 +|D_-|^2}}\right]
	 ,\nonumber\\
	 \label{G}
   &	 G \equiv \frac{(\omega_0 x_z)^2\gamma \left[|D_+|^2 +|D_-|^2\right]}{L^2\left(\gamma^2 + \Omega^2\right)}\,,
	 \\
  & d_\text{th} \equiv \frac{D_-b_\text{th} +D_+ b^\dag_\text{th--}}{\sqrt{|D_+|^2 +|D_-|^2}} \,,\quad
	 e^{2i\eta}\equiv \frac{\gamma+i\Omega }{\gamma-i\Omega }\,.
\end{align}
\end{subequations}
The term \eqref{aout1} describes the optical shot noise and $d_\text{th}$ stands for the thermal noise of the mechanical oscillator upconverted to the optical domain, in last signal term we should take in mind that $f_s(\Omega)=-f_s^*(-\Omega)$ in accordance with definition \eqref{fs} and since $F_s(t)$ is a real. As can be verified by direct calculations, complex amplitude $a^\dag_\text{in--}$ is absent in \eqref{aout1} (for details see Appendix~\ref{BA}).

Let us discuss the difference between the force detection techniques that involve either a monochromatic probe light (conventional one) or the dichromatic light (considered in this paper). In the first case the state of the light leaving the system, $a_{out}$, is distorted by the interaction with the mechanical degree of freedom. The light becomes squeezed. The degree of squeezing depends on the optical power and on the spectral frequency. A variational technique is needed for an accurate detection of the mechanical force using $a_{out}$. In the second case, the output light is in the coherent state for the symmetric pump $|D_+| =|D_-|$ (as well as $|A_+|=|A_-|$). Both the fluctuational,  \eqref{initbx2} and dynamic, \eqref{tildegamma} terms do not contribute to the quantum state of the output field. There is no power dependence of $a_{out}$ (in the case of zero mechanical attenuation, $\gamma_m=0$). The fluctuation of the output light at the resonant optical frequency does not depend on the light induced 
fluctuations of the coordinate as well as momentum of the movable mirror. This is ideal for the force detection and is one of the major finding of present study.

We measure not the coordinate of the mechanical oscillator, but its quadrature. The phases associated with the light amplitude $D_\pm$ (let’s label them $\varphi_\pm$, so that $D_\pm=|D_\pm|e^{i\varphi_\pm}$) are crucial for the presented setup. The relative phase $\varphi_r=(\varphi_+-\varphi_-)/2$ of $D_+$ and $D_-$ defines which particular mechanical quadrature is measured. One can select phases of $D_+$ and $D_-$ so that combination $C=D_-f_s + D_+f_{s-}^*$ in \eqref{aout1} is zero. It corresponds to the case when we measure mechanical quadrature which is not excited by signal force. In contrast, if the phases are selected so that combination $|C|$ reaches maximal possible value, we measure exactly the quadrature excited by the signal force. It is this idea of QND measurement of the mechanical quadrature was proposed in \cite{80a1BrThVo}.

\subsection{Detection}\label{Detection}

In what follows we discuss the detection procedure of the force using the output light. The direct amplitude measurement does not provide any information on the mechanical degree of freedom. An optimized processing of the output light is needed to retrieve the information. For instance, synodyne detection technique can be utilized.  The technique involves the bichromatic optical pump as the LO. The LO is phase shifted (delayed) by an angle $\theta$ as shown on  Fig.~\ref{Dpump2} (compare with \eqref{mathA}):
\begin{align}
 \mathcal A(t) = e^{-i\theta}\left( A_+e^{-i(\omega_0 +\omega_m) t} + A_-e^{-i(\omega_0-\omega_m) t }\right)
\end{align}
The LO is mixed with the output light and is detected using the balanced homodyne scheme. The differential current $I_-$ is proportional to
  \begin{align}
    I_- & \sim \mathcal A(t) a^\dag_\text{out}(t) e^{i\omega_0 t}
	+ \mathcal A^*(t) a_\text{out}(t) e^{-i\omega_0 t}
  \end{align}
In the case of symmetric pumps  $|A_+|=|A_-|$ the current amplitude can be presented in form
\begin{align}
\label{I-}
 I_- &\sim \cos (\omega_m t +\phi_r) \left[ e^{i \theta}  a_\text{out}(t) + e^{-i\theta} a^\dag_\text{out}(t) \right ],
\end{align}
where $\phi_r$ (different from $\varphi_r$) is the relative phase of the pump light harmonics $A_\pm$. See details in Appendix~\ref{app}.

Presenting the differential current as a sum of the signal, $I_s$, and noise, $S_I$, parts, where the last one is characterized with the single-sided spectral density $S_I$, we derive expressions
\begin{align}
\label{Is}
 I_s (\Omega) &= \frac{\sqrt {2G}\,e^{i\eta}\sin (\theta-\phi_r)}{\gamma_m -i(\Omega-\omega_m)}\, \mathsf f_\phi ,\\
 \mathsf f_\phi  &\equiv  e^{i(\beta -\phi_r)}f_s^*(\omega_m-\Omega ) + e^{-i(\beta -\phi_r)}f_s(\Omega -\omega_m) ,\nonumber\\
  & e^{2i\beta} \equiv \frac{\gamma+i\omega_m }{\gamma-i\omega_m } ,\\
 \label{SI}
 S_I(\Omega) &= 2 + \frac{4G\gamma_m\left(2n_\text{th}+1\right)\sin^2(\theta-\phi_r)}{\gamma_m^2 + (\Omega-\omega_m)^2},
\end{align}
where the correlators and commutators (\ref{comm1}, \ref{comm2}) for $a_\text{in}$ and $b_\text{th}$ were utilized.

We can tune the measured quadrature $\mathsf f_\phi$ of the signal force varying phase $\phi_r$. The first term in Eq.~\eqref{SI} results from the optical shot noise. This term is doubled if compared with the case of the force measurement using the single pump harmonic because of the sidebands at frequencies $\omega_m \pm \Omega$. The second term describes the thermal fluctuations of the mechanical oscillator. The spectral density of the current is constant in the case of zero mechanical loss ($\gamma_m=0$) indicating absence of the back action quantum noise. As we have mentioned, $I_s (\Omega)$ does not carry any information about the mechanical degree of freedom for $\theta-\phi_r=0$, which corresponds to amplitude detection.

It is convenient to rewrite Eqs.~(\ref{Is}, \ref{SI}) as spectral density $S_{\mathsf f}$ of normalized force $\mathsf f$:
\begin{align}
  \label{Sf}
 S_{\mathsf f}(\Omega) &= \frac{\gamma_m^2 +(\omega_m-\Omega)^2}{G\sin^2(\theta-\phi_r)}
      + 2\gamma_m \left(2n_\text{th}+1\right).
\end{align}
The sensitivity of the measurement is limited by the second term originating from the thermal fluctuations of the mechanical oscillator. The first term describes measurement error due to the optical shot noise and  decreases as the pump power increases. The minimal detectable amplitude $F_s$ of resonance force with duration $t_F$  can be much smaller than the SQL of the force measurement $F_{SQL}= 2\sqrt{\hslash m\omega_m}/t_F$ \cite{Braginsky68,BrKh92}.

Indeed, for the case of the vanishing mechanical losses ($\gamma_m \to 0$) and optimized delay phase $\theta$
($\sin (\theta- \phi_r)=1$) we integrate spectral density \eqref{Sf} inside bandwidth $\Delta \Omega \simeq 2\pi/t_F$ around  frequency $\omega_m$  to obtain the measurement sensitivity of $F_s$
\begin{align}
\label{Fs}
\left|\frac{F_s}{
\sqrt{2\hslash m\omega_m} }
\right|^2 > \int_{\Delta \Omega} S_{\mathsf f}(\Omega)\, \frac{d\Omega}{2\pi}
  ,\ \Rightarrow\
 F_s > \frac{\pi\sqrt 2}{\sqrt{3Gt_F}} \cdot F_{SQL},
\end{align}
We see that the detectable force is smaller than $F_{SQL}$ at large enough pump power $Gt_F\gg 1$.

Let us compare the measurement strategy proposed in this paper with the conventional variational measurement strategy \cite{93a1VyMaJETP, 95a1VyZuPLA, 02a1KiLeMaThVy}. The back action avoiding is feasible in the conventional variational scheme only in a narrow spectral frequency band specific for a properly selected quadrature amplitude. In contrast, proposed here measurement technique is not hindered by back action. Since the symmetric dichromatic pumping is essential for the back action evading we call the described here technique as a synodyne variational measurement. The complete subtraction of back action occurs because the susceptibility $\chi$ of the mechanical system is imaginary at the resonance, $ \chi(-\omega_m) =-\chi(\omega_m)$. Recall, the same property is used in the synodyne detection \cite{
Buchmann2016}.

The dichromatic pump is essential for the measurements while a variety of local oscillators can be used to retrieve the useful information from the light leaving the system. A monochromatic local oscillator can be used instead of the dichromatic one. There is no need in a special preparation of the local oscillator to achieve back action evading measurement, as was done in the variational measurement. This fact significantly simplifies the measurement procedure.

Mechanical loss destroys the back action avoiding balance. The amplified fluctuations of the mechanical thermal bath, penetrating the system due to presence of finite mechanical attenuation, upconverted into the optical domain behave as standard optical back action terms. They increase with the optical power. This type of unconventional back action confirms the {\em parametric} nature of the process. In contrast, in the case of the conventional coordinate measurement the back action results in the excitation of the mechanical coordinate by light that leads to the dependence of the phase of the output light on the amplitude of the input light. The thermal mechanical fluctuations are not amplified in this case.

\section{Corrections due to asymmetric pump}

We have considered the completely symmetric opto-mechanical system. Let us analyze briefly more realistic situations in what follows. In the case of asymmetric but resonant pump $|D_-| \ne |D_+|$ the back action terms do not disappear and the system may become unstable. It also may become unstable if the center of the optical doublet is shifted with respect of the optical mode.  In the case of nonresonant pump, $\delta =\omega_+-\omega_0 -\omega_m=\omega_0-\omega_- -\omega_m\ne 0$ (where $\omega_\pm$ are the frequencies of the pump harmonics), the system is stable and dynamic back action is still zero, however, the fluctuation back action may be incompletely compensated.

In general, all these asymmetry effects are of technical nature and can be prevented using classical feedback keeping the system in the stable symmetric state. There exist another effect that cannot be prevented. We studied the resolved sidebands case and neglected the contribution of the high order optical harmonics generated in the system by the mirror vibrating at the doubled mechanical frequency $2\omega_m$. The first two harmonics of this type are detuned  from the resonance frequency $\omega_0$ by approximately $\pm 2\omega_m$. Taking them into account results in the following modification of Eq.~\eqref{Sf} (see details in Appendix \ref{app2}):
\begin{align}
  \label{Sf2}
 S_{\mathsf  f}(\Omega) &= \frac{\gamma_m^2 +(\omega_m-\Omega)^2}{G\sin^2(\theta-\phi_r)}
      + 2\gamma_m \left(2n_\text{th}+1\right) + \nonumber\\
     &\qquad  \frac{G  }{\omega_m^2} \left(\gamma^2 +(\Omega-\omega_m)^2\right)\, ,
\end{align}
where the last term represents the residual quantum back action noise.

Presence of the noise limits the force measurement sensitivity. Indeed, for the case of $\gamma_m=0$,
$\sin(\theta - \phi_r)=1$, and $\gamma t_F\gg 1$ we find
\begin{align}
 F_s >  \sqrt\frac{2\pi\, \gamma}{\sqrt 3\, \omega_m}
 \cdot F_{SQL},
\end{align}
which is achieved at the optimal pump power
\begin{equation} \label{gopt}
G_{opt} \simeq \frac{\omega_m }{\gamma t_F}
\end{equation}
The detectable force $F_s$ can be much less than the SQL since $\gamma\ll \omega_m$.

\section{ Dynamic stability analysis}

We truncated non-resonance terms in the considered Hamiltonian (\ref{Hamiltonian}) and initial equations \eqref{initEq}. Let us focus at the regular pondermotive force proportional to $\sim \mathcal D \mathcal D^*$ in \eqref{initEq2}. It contains a constant term, which can be compensated in an experiment by an external feedback, and oscillating terms $\sim D_+ D_-^*e^{- 2i\omega_m t} + D_+^* D_-e^{ 2i\omega_m t}$. The last ones are non-resonant. However, they increase with power and may become large, so we have to take its into account and rewrite \eqref{bm} in form
\begin{align}
  \hat b(t) & = \mathcal B(t)
  + e^{-i\omega_m t}\int_{-\omega_m}^\infty b (\Omega) \, e^{-i\Omega t} \,  \frac{d\Omega}{2\pi},\\
  \mathcal B(t)&  = \frac{\omega_0x_z}{3\omega_m L}\left(D_+D_-^* e^{-2i\omega_mt}
    + D_+D_-^* e^{2i\omega_mt}\right)
\end{align}
After substitution terms $\mathcal B(t),\ \mathcal B^*(t)$ into \eqref{initEq1} one can find nonlinear modifications to the regular amplitudes, so instead of $D_\pm$ \eqref{Dpm} we get new amplitudes $\tilde D_\pm$ defined as
\begin{align}
 \tilde D_+ &= D_+\left(1 + \frac{4\omega_0x_z^2}{3\omega_m L^2}  |D_-|^2\right),\\
 \tilde D_- &= D_- \left(1 - \frac{4\omega_0x_z^2}{3\omega_m L^2} |D_+|^2 \right)
\end{align}
It is easy to see that amplitude $\tilde D_+$ slightly increases whereas $\tilde D_-$  decreases. This  is the origin of additional {\em negative} damping $\gamma_m^\text{add}$  introduced into equations for amplitudes $b,\ b^\dag$ of mechanical oscillator  even for the case of pure symmetric pump ($|A_+| = |A_-|$). We obtain
\begin{align}
\dot b  + \left ( \gamma_m - \gamma_m^\text{add} \right ) b & = 0, \\
  \gamma_m^\text{add} & = \frac{G^2 \gamma}{3\omega_m^2}.
\end{align}
This negative damping increases non-linearly with increase of pump $G$ so that the system may become unstable at a certain pump power. It is especially important for the ideal case when $\gamma_m$ is small. In this case $ \gamma_m^\text{add} \approx (\gamma t_F^2)^{-1}$ (see Eq.~\ref{gopt}). This number can be significant as compare with $\gamma_m$. For instance, if $t_F=100/\gamma$, $\gamma_m=10^{-4} \gamma$, and $\omega_m=10 \gamma$, this condition is certainly fulfilled. For example, for parameters  $G\simeq 2\cdot 10^5\, \text{s}^{-1},\ \gamma \simeq 10^6\,\text{s}^{-1},\ \omega_m \simeq 3\cdot 10^7\,\text{s}^{-1}$, used in \cite{Clerk2016}, we obtain $\gamma_m^\text{add}\simeq 13\,\text{s}^{-1}$ which is comparable with  $\gamma_m\simeq 24\,\text{s}^{-1}$ observed in experiment.

The negative damping occurs because the dichromatic optical pump excites the mechanical harmonic $x_{\pm 2}$ at the frequency separation of the pump harmonics $2\omega_m$.  These off-resonant mechanical oscillation at the doubled mechanical frequency  modulates the eigen frequency of the FP cavity, which in turn changes the effective amplitudes $D_\pm $ of the pump in the cavity.  A detailed analysis shows that amplitude $D_-$ effectively decreases, while $D_+$ -- increases. The balance between the positive mechanical damping introduced by $D_-$ and negative one produced by $D_+$ becomes compromised. It is the reason why negative damping $\gamma_m^\text{add} \sim |D_+|^2 -|D_-|^2$ appears.

It is possible to suppress negative damping $\gamma_m^\text{add}$ simply by reducing the ponderomotively forced second mechanical harmonics $x_{\pm 2}$. Application to the mirror of a classical periodic force that has the same magnitude as the ponderomotive one and phase shifted by $\pi$ would suppress the oscillation and removes the instability. Another way of removal of the instability is an introduction of a controlled imbalance (i.e. $|A_-|$ slightly larger than $|A_+|$) that results in a residual ponderomotive attenuation.

Therefore, the instability is not a fundamental limitation of the measurement technique. On the other hand, the importance of the negative damping we found goes beyond the force measurement problem, as we found that the opto-mechanical system may become unstable even though the opto-mechanical gain is compensated by the opto-mechanical loss. This is a novel type of an opto-mechanical oscillator that worth an additional study.

\section{Conclusion}

We have shown that a back action avoiding measurement of a classical resonant force acting on a movable mirror being a part of a Fabry-Perot optical cavity can be realized by involvement of a dichromatic probe light. The probe harmonics have the same power and are detuned by the same absolute frequency from the optical resonance in a way that the dynamic back action completely cancels for the optical resonant spectral frequencies. While the system becomes dynamically unstable even in the absolutely symmetric case, the instability can be suppressed by usage of a classical force compensating for the classical ponderomotive force impinged by the optical pump on the movable mirror of the cavity. The signal at these frequencies can be detected by means of a dichromatic homodyne (synodyne) detection. The method is advantageous for advanced metrology experiments, involving mechanical force resonant with the suspended mirror.

\section*{Acknowledgments}

Authors are very grateful to Mikhail Korobko for fruitful discussions. Sergey P. Vyatchanin acknowledges support from the Russian Foundation for Basic Research (researches on Sec.~IV supported by grant No. 16-52-10069), Russian Science Foundation (researches in Sec.~II supported by grant No. 17-12-01095) and National Science Foundation (researches on Sec.~III supported by  grant No. PHY-130586).

\appendix

\section{Back action evading at the output}\label{BA}

Here we provides detailed derivation of \eqref{aout1} to explain why back action is absent. First we write down combination $B=\left[ D_- b + D_+ b^\dag_-\right]$ in \eqref{aout} using \eqref{initbx} and  keeping only back action terms \eqref{initbx2}:
\begin{align}
 B_{ba} &= \frac{2 k x_0 \sqrt{\gamma/\tau}\big[|D_+|^2 - |D_-|^2\big]a_\text{in} }{i(\gamma -i\Omega)( \gamma_m + \Gamma -i\Omega)}+\\
      &\qquad + \frac{2 k x_0 \sqrt{\gamma/\tau}\big[D_+D_- - D_-D_+\big]a^\dag_\text{in--} }{i(\gamma -i\Omega)( \gamma_m + \Gamma -i\Omega)}
\end{align}
Underline, term $\sim a^\dag_\text{in--}$ is absent here.
After substitution into \eqref{aout} we have (keeping only back action terms):
\begin{align}
 a_\text{out}|_\text{ba} &= \frac{\gamma+i\Omega }{\gamma-i\Omega }\cdot a_\text{in} +\\
	    &\qquad +\frac{\big(2kx_0\big)^2\gamma\left[|D_+|^2 -|D_-|^2\right]}{ \tau
	    \big(\gamma-i\Omega\big)^2\big( \gamma_m + \Gamma -i\Omega\big)}\cdot a_\text{in}
\end{align}
Here first term describes shot noise, second one --- back action. Finally, using definition \eqref{tildegamma} one can compact these terms in form \eqref{aout1} (term $\sim a^\dag_\text{in--}$ is absent) .

\section{Synodyne detection}\label{app}

Here we consider in detail formulas in Sec.~\ref{Detection}. The LO is mixed with the output light and detected using the balanced homodyne scheme, so differential current $I_-$ is equal to
  \begin{align}
    I_- & = K\left( \mathcal A(t) a^\dag_\text{out}(t) e^{i\omega_0 t}
	+ \mathcal A^*(t) a_\text{out}(t) e^{-i\omega_0 t}\right) =\\
	&= K \left[ a^\dag_\text{out}(t)e^{-i\theta}\left(A_+e^{-i\omega_mt} +A_-e^{i\omega_m t}\right) +\right. \\
	&\quad +\left. a_\text{out}(t)e^{i\theta}\left(A_+^*e^{i\omega_mt} +A_-^*e^{-i\omega_m t}\right)\right]
  \end{align}
Where $K$ is a constant. For symmetric pump $|A_+|=|A_-|$ we obtain, introducing phases $A_+ =|A_+|e^{i\phi_+},\ A_-=|A_-|e^{i\phi_-}$ and putting $K=1/(\sqrt 2 |A_+|)$:
\begin{align}
\label{I-b}
 I_-& =\sqrt 2  \cos(\omega_m t+\phi_r)\times\\
    &\quad \times \left[e^{-i\theta+i\phi_s}a^\dag_\text{out}(t)
     + e^{i\theta-i\phi_s}a_\text{out}(t)\right] ,\nonumber\\
    &\phi_s =\frac{\phi_- + \phi_+}{2},\quad \phi_r =\frac{\phi_- - \phi_+}{2}
 \end{align}
Below we put sum phase to zero, $\phi_s=0$.
We can rewrite \eqref{I-b} in spectral domain:
\begin{align}
  \label{I-Om}
 I_-(\Omega)& =  \frac{e^{i\theta}}{\sqrt 2}\left\{a_\text{out}(\Omega +\omega_m) e^{i\phi_r}
    + a_\text{out}(\Omega -\omega_m)e^{-i\phi_r}\right\} +\\
    + \frac{e^{- i\theta}}{\sqrt 2} &\left\{a^\dag_\text{out}(-\Omega +\omega_m) e^{i\phi_r}
    + a^\dag_\text{out}(-\Omega -\omega_m)e^{-i\phi_r}\right\}.\nonumber
\end{align}

We see that signal bandwidths are around frequencies $\pm\omega_m$. Substituting \eqref{aout1} into \eqref{I-b} we get signal part \eqref{Is} of differential current at frequencies $\Omega \simeq \omega_m$. Calculating single-sided density $S_I$ \eqref{SI}, which is proportional to symmetrized product $\big[I(\Omega) I^\dag (\Omega) +  I^\dag (\Omega)I(\Omega)\big]$, we have to account additional contribution of vacuum optical fluctuations detuned from cavity resonance by $\pm 2\omega_m$ -- that is why first term $(2)$ appears. The second term in \eqref{SI} is originated by terms proportional to $d_\text{th}$ in \eqref{aout1}.

\section{To derivation of (\ref{Sf2})}\label{app2}

We have to account in ponderomotive force the terms containing non-resonance terms, i.e. we have to account in \eqref{initEqFT2}
additional regular but non-resonance term $x_{add}(t)$ in right part:
\begin{align}
  \label{initEqFT4}
  b_{add} &(\gamma_m-i\Omega) = \frac{i\omega_0 x_z}{L}\, \left[D_+^* d_{2\omega_m} + D_- d^\dag_{-2\omega_m}\right],\\
  d_{2\omega_m} &\equiv d_i(2\omega_m + \Omega),\quad d^\dag_{-2\omega_m}\equiv d_i^\dag(-2\omega_m -\Omega)
\end{align}
optical components $d_{2\omega_m}, \ d^\dag_{-2\omega_m}$ are out of resonance and for them we have good approximation:
\begin{align}
  d_{2\omega_m} &\simeq \frac{\sqrt{2 \gamma}\,a_{2\omega_m}}{-2i\omega_m},\quad
    d^\dag_{2\omega_m} \simeq \frac{\sqrt{2 \gamma}\,a^\dag_{2\omega_m}}{2i\omega_m},\\
    a_{2\omega_m} &\equiv a_{in}(2\omega_m + \Omega),\ a^\dag_{-2\omega_m}\equiv a_{in}^\dag(-2\omega_m -\Omega).
\end{align}
Then we have to calculate combination $(D_-b_{add} + D_+b^\dag_{add-})$ and substitute as additional term into \eqref{aout}
\begin{align}
  a^\text{add}_\text{out} &\simeq
    - \frac{\omega_0 x_z}{\omega_m L}\cdot \frac{\gamma}{(\gamma-i\Omega)(\gamma_m -i\Omega)}\times\\
  &\quad \times \left\{D_-\left[D_+^*a_{2\omega_m} +D_-a^\dag_{-2\omega_m}\right] \right.+\\
  &\qquad \left. + D_+\left[D_+a^\dag_{2\omega_m-} +D_-^*a_{-2\omega_m-}\right] \right\},\\
  a_{-2\omega_m-} &\equiv a(-2\omega_m +\Omega),\quad a^\dag_{2\omega_m-}= a^\dag(-2\omega_m + \Omega).\nonumber
\end{align}
Now substituting $a^\text{add}_\text{out}$ into \eqref{I-Om} one can calculate additional term in formula \eqref{SI} for spectral density $S_I$ which can be recalculated into last term of $S_{\mathsf f}$ in \eqref{Sf2}.



\end{document}